\documentclass[10pt, conference, compsocconf]{IEEEtran}
\ifCLASSINFOpdf
\else
\fi
\usepackage[cmex10]{amsmath}
\usepackage{algorithmic}

\usepackage{array}

\usepackage{mdwmath}
\usepackage{mdwtab}
\usepackage{url}

\usepackage{balance}  
\usepackage{graphicx} 

\begin{document}
%
\title{Effective Blog Pages Extractor for Better UGC Accessing}


\author{
\IEEEauthorblockN{Kui Zhao\IEEEauthorrefmark{1},
Yi Wang\IEEEauthorrefmark{1},
Xia Hu\IEEEauthorrefmark{2},
Can Wang\IEEEauthorrefmark{1}
}
\IEEEauthorblockA{\IEEEauthorrefmark{1}
Zhejiang Provincial Key Laboratory of Service Robot\\
College of Computer Science and Technology\\
Zhejiang University\\
Hangzhou, China, 310027\\
\{zhaokui,wangyi,wcan\}@zju.edu.cn}

\IEEEauthorblockA{\IEEEauthorrefmark{2}Hangzhou S\&T Information Research Institute\\
Hangzhou, China, 310014\\
hx@hznet.com.cn}
}


%


\maketitle

\begin{abstract}
Blog is becoming an increasingly popular media for information publishing. Besides the main content, most of blog pages nowadays also contain noisy information such as advertisements etc. Removing these unrelated elements can improves user experience, but also can better adapt the content to various devices such as mobile phones. Though template-based extractors are  highly accurate, they may incur expensive cost in that a large number of template need to be developed and they will fail once the template is updated. To address these issues, we present a novel template-independent content extractor for blog pages. First, we convert a blog page into a DOM-Tree, where all elements including the title and body blocks in a page correspond to subtrees. Then we construct subtree candidate set for the title and the body blocks respectively, and extract both spatial and content features for elements contained in the subtree. SVM classifiers for the title and the body blocks are trained using these features. Finally, the classifiers are used to extract the main content from blog pages. We test our extractor on 2,250 blog pages crawled from nine blog sites with obviously different styles and templates. Experimental results verify the effectiveness of our extractor.

\end{abstract}

\begin{IEEEkeywords}
UGC; Blog; Content Extraction; DOM-Tree;

\end{IEEEkeywords}

%
\IEEEpeerreviewmaketitle

\section{Introduction}
As a platform for people to publish contents they are interested in, 
blog is becoming an important part of online culture. 
In fact,  the number of blog users has been increasing  rapidly. Taking China as an example, 
according to the survey published by CNNIC, 
the number of blog users in China is 109 million in 2014 with a 21 million increase from the user number in 2013. 
\footnote{http://www.cnnic.net.cn/hlwfzyj/hlwxzbg/, in Chinese.}

\begin{figure}[ht]
\centering
\includegraphics[height=6cm]{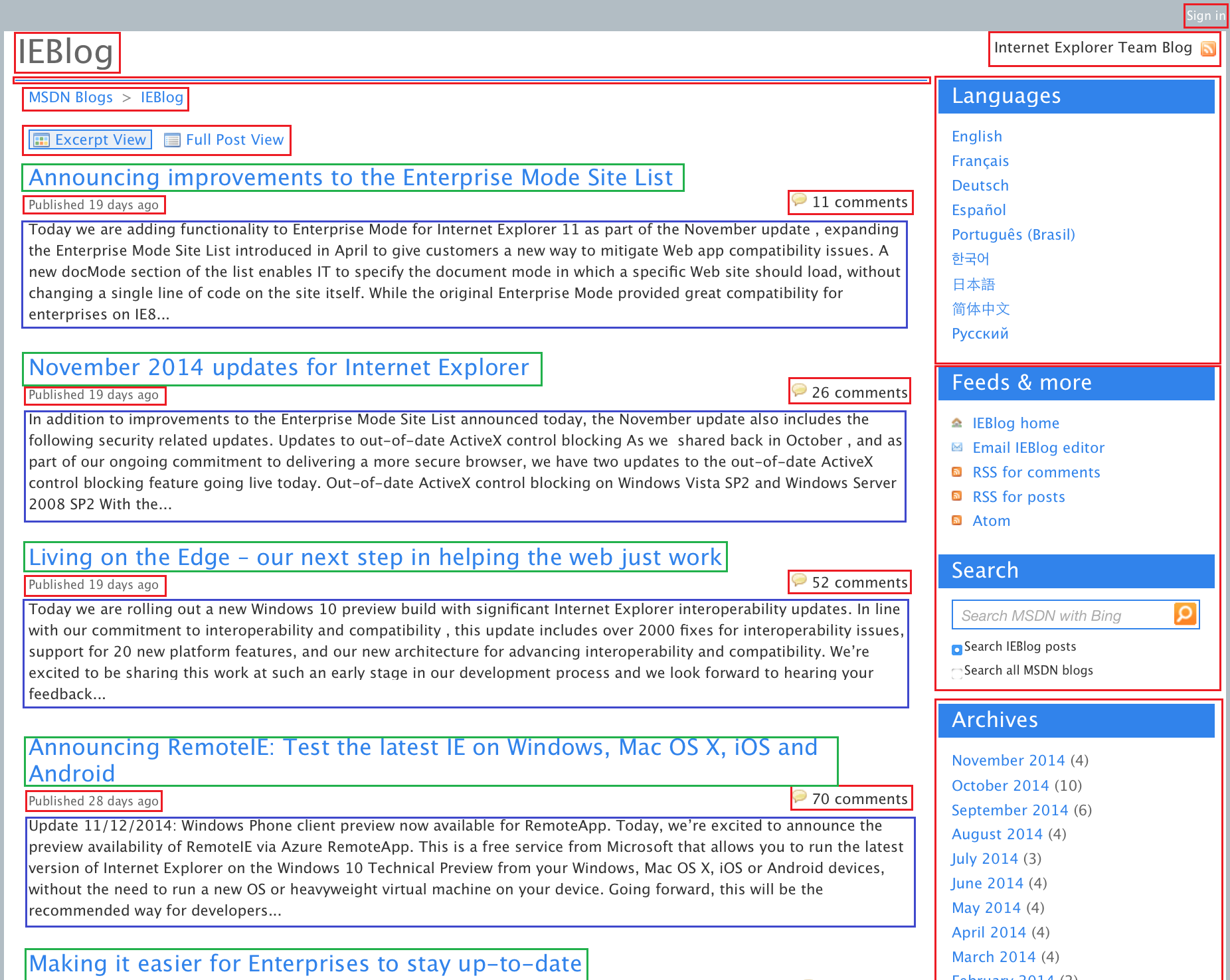}
\caption{\label{fig:blog-home-page}
A typical blog page, 
with irrelevant contents marked by red boxes, and main contents including titles and bodies marked in green boxes and blue boxes respectively. 
}
\end{figure}

However, as shown by the example in Fig.\ref{fig:blog-home-page}, 
most blog pages contain irrelevant contents such as pop-up
ads, decorative images, navigation links etc. \footnote{http://blogs.msdn.com/b/ie/.}
Eliminating these unrelated contents can help users better browse information they
are interested in, especially for people with visual impairment, who access web contents via screen readers. Significant improvements in accessibility can be made if the extracted contents instead of the raw HTML is fed to screen readers. What's more, it is much easier to adapt the extracted content to diversified terminals such as mobile phones, PDAs etc. Recent years have witnessed many successful applications using the extracted content to better serve their users, such as KnowLife \cite{ernst2014knowlife}, Odin \cite{hailpern2014odin} etc. 
These applications need to extract, store and process the useful information from webpages by removing irrelevant information.  As blog is gaining popularity and becoming a major source of User Generated Contents(UGC), extracting contents from blog pages becomes an important research topic.

Most of the traditional webpage content extraction methods are template-based, such as \cite{buyukkokten2001accordion}, 
\cite{chen2006template} and many others.  Although these template-based methods are highly accurate, they may incur expensive cost in that different extractors need to be developed for webpages of different templates. Moreover, template-based extractors will fail once the template of a site is updated. As blog sites nowadays are using greatly diversified templates, traditional extraction methods are expensive and inconvenient solutions. Moreover, many blog sites provide users with the capability to configure templates, 
which makes template-based solutions more difficult. 

To overcome the limitation of template-based extraction, many researchers in recent years attempt to develop 
template-independent extractors for webpages. Wang et.al developed a template-independent extractor for news 
pages by first identifying the title block and then extracting the body block\cite{wang2009can}. 
Sun et. al extracted the webpage content by exploiting the text density in DOM-Tree nodes \cite{sun2011dom}. 
These extractors either only work on webpages with simple layout  \cite{wang2009can}, 
or only target at the main body of the webpage content \cite{sun2011dom}. 
In contrast, effectively extracting contents from blog pages is a challenging issue in that:

(1) Blog pages have more complex structures than news pages;

(2) Multiple titles and bodies might be contained in a single blog page;

(3) Blog posts are usually written in a less rigorous way than news articles, 
thus the relevance between the title and the body in blog pages 
is less certain than that in news pages.

In this paper, we propose a novel method to extract titles and bodies from blog pages in a template-independent way. 
By first parsing a blog page into a DOM-Tree, all elements including titles and bodies in this page are represented by subtrees. 
Before extracting subtrees corresponding to titles and bodies using supervised learning, 
two subtree candidate sets are constructed, one for titles and another for bodies. 
Then the training set is formed by manually labelling the title subtrees and body subtrees, both spatial features and content 
features are extracted from the elements contained in subtrees. 
These features are first used to train the SVM classifier for title subtrees. 
The title subtrees are further used to construct new features for identifying body subtrees. 
Encouraging experimental results are achieved on 2,250 blog pages from nine very different blog sites.

The rest of the paper is organized as follows: 
we briefly review related work in section 2 
and describe our method with details in section 3 and 4. 
Then we show our experimental setup and results, 
and discuss the results in section 5. Finally conclusions come out and 
we plan the future research in section 6.

\section{Related Work}
Early work in content extraction from HTML documents can be traced back to \cite{rahman2001content} by Rahman et al. In last decade, with the explosive growth of the information on the web, content extraction from webpages has attracted attention from research and industrial communities. 

Among all methods, template-based ones are dominant because they are effective and easy to implement. 
The most simplistic ones are handcrafted web scrapers, 
including NoDoSE \cite{adelberg1998nodose} and XWRAP \cite{buyukkokten2001accordion}. 
They extracted contents from webpages  with a common template by looking for special HTML cues using regular expressions.
A different category of template-based methods are  using {\it Template Detection}, 
such as Bar-Yossef's \cite{bar2002template}, Chen's \cite{chen2006template} and  Yi's \cite{yi2003eliminating}, 
in which webpages with the same template are used to learn the common structure. 
These methods will first identify the template and then extract contents by removing all parts found in every webpage with the same template. The major problem with template-based extractors is that different extractors must be developed for different templates, making the solution expensive. What's more, once the template is updated, as happens frequently in many web sites, the corresponding extractor will be invalidated.

To overcome the limitation of the template-based extractors, many researchers attempted to extract webpage content in a template-independent way in recent years.
Cai et al. \cite{cai2003vips} proposed a vision-based page segmentation algorithm named VIPS, extracting the semantic structure of a webpage based on its DOM-Tree and visual presentation. 
Song et al. \cite{song2004learning} thought the rule-based method in VIPS was too simple and developed a model that assigns importance scores to blocks in a webpage according to their spatial features and content features. 
Zheng et al. \cite{zheng2007template} proposed a template-independent news page extraction method based on visual consistency. Wang et al. exploited more features and achieved a higher accuracy in the news title and body extraction \cite{wang2009can}. 
In their method, a template-independent extractor was developed for news pages by first identifying the title block 
and then extracting the body block. Some more recent methods exploited statistical information on webpages, such as 
CETR \cite{weninger2010cetr} and CETD\cite{sun2011dom}. 
These methods extract contents by identifying the regions with high text density, 
i.e., regions including many words and few tags are more likely to be contents. 
Qureshi et al. proposed a hybrid model \cite{qureshi2012hybrid} by considering not only statistical information but also formatting characteristics. 

Although the above methods have achieved some success in extracting webpage content in template-independent ways, 
developing template-independent extractors for blog pages still poses particular challenges not well addressed in any previous 
work. The complicated page structure, existence of multiple titles and bodies, random style of writing etc. 
make the template-independent extraction for blog pages a difficult task.

\section{Content Extraction}
\subsection{DOM-Tree}
\begin{figure}[ht!]
\centering
\includegraphics[height=9cm]{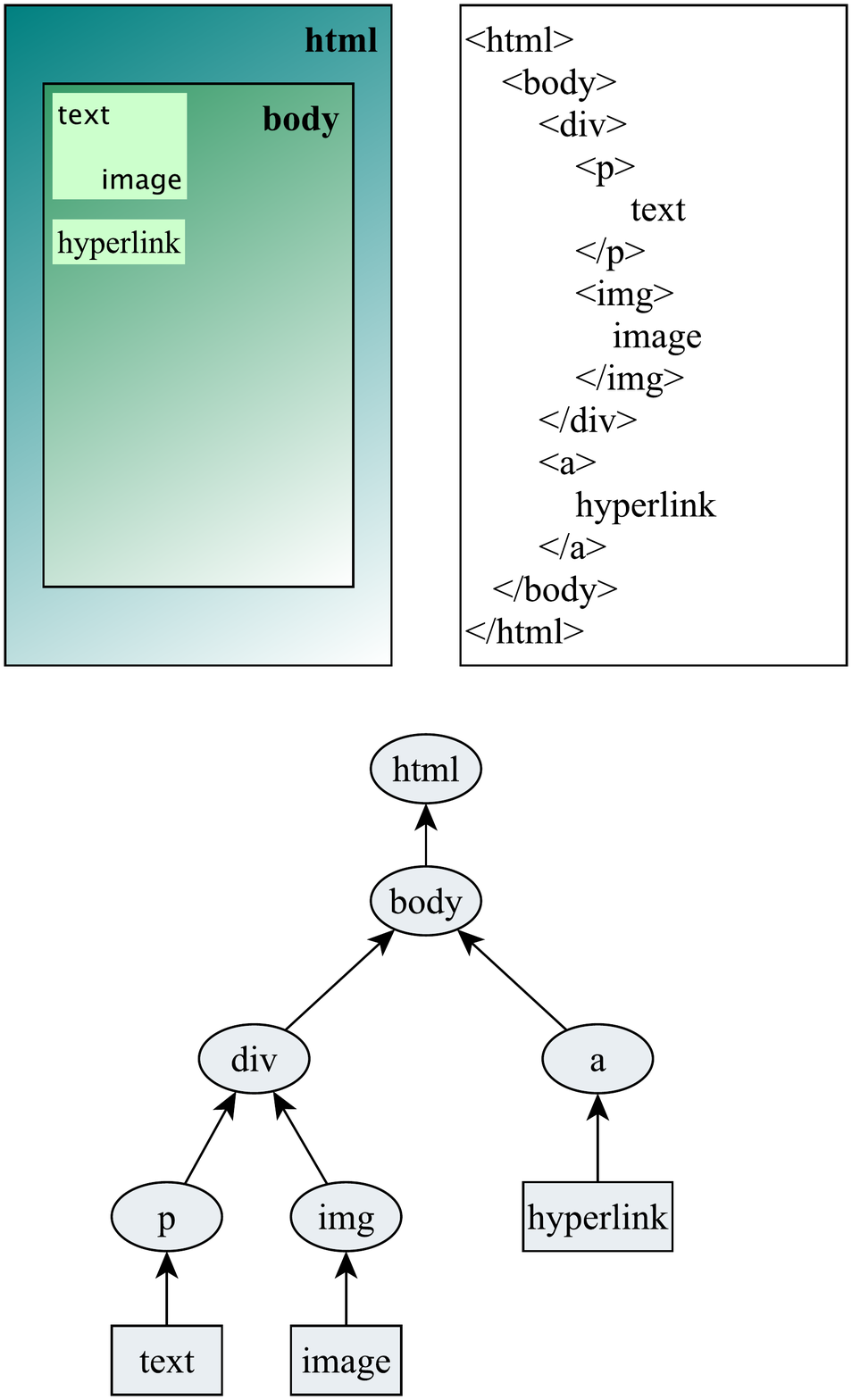}
\caption{\label{fig:dom}
An example of DOM-Tree, on the right side is the DOM-Tree for the HTML code on the left side.}
\end{figure} 

We use $\bf{P}$ to denote the blog page aimed to extract titles and bodies. 
We use $\bf T_{DOM}$ to denote the DOM-Tree(an example shown in Fig.\ref{fig:dom}) built from $\bf P$. 
We use $\bf T$ to denote the set of subtrees in $\bf T_{DOM}$. 

A subtree $\bf{s}_i\in T$ is called a \textit{title subtree} if
${\bf s}_i$ but not any subtree of ${\bf s}_i$ contains the title. 
Similar definition is made for a \textit{body subtree}
$\bf b_i \in T$. Our purpose is to
find out the set of title subtrees $\bf S\subset T$ and the
set of body subtrees $\bf B\subset T$. 
More precisely, given a subtree $\bf{t}_i\in T$, our task is to tell whether $\mathbf{t}_i$ is in the set $\bf S$ or in the set $\bf B$ 
or neither.
\subsection{Extraction Process}
\begin{figure}[ht]
\centering
\includegraphics[height=1.9cm]{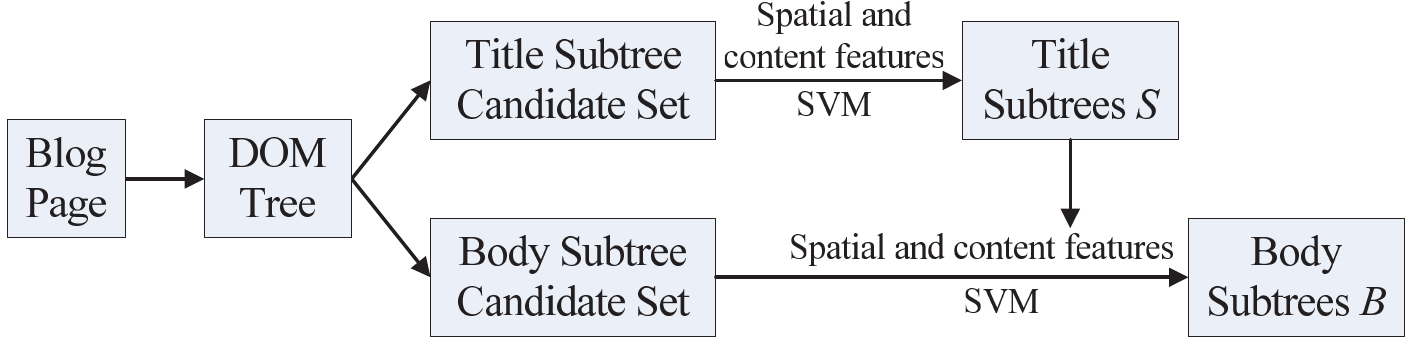}
\caption{\label{fig:process}The extraction process of blog titles and bodies.}
\end{figure}

We parse a blog page into a DOM-Tree and try to find title and body subtrees. 
In order to improve the efficiency of extraction, the subtree candidate set is generated
for title and body subtrees respectively. Then novel features described 
in section 4 are constructed and we use them to learn extraction models. 
As titles in a blog page usually share the same style and bodies are always sandwiched 
between two titles in the vertical direction, title subtrees are extracted
first and then used to help construct new features for identifying body
subtrees. 
Finally, we use SVM classifiers \cite{burges1998tutorial} with 
Gaussian RBF kernel \cite{chapelle1999support} to identify
title and body subtrees in their subtree candidate sets respectively. 
The whole process is illustrated in Fig.\ref{fig:process}.

\subsection{Candidate Set}
A DOM-Tree always contains thousands of nodes. 
To make it more efficient to extract titles and bodies, 
we construct the candidate set for title and body subtrees by selecting proper subtrees respectively: 

\subsubsection{Title Subtree Candidate.} 
All titles have some textual contents, so we choose subtrees containing 
at least one textual leaf node as title subtree candidates. 

\subsubsection{Body Subtree Candidate.} 
Unlike titles, bodies may contain only images without any textual contents. 
So the rule for selecting title subtree candidates cannot be applied to selecting body subtree candidates. 
By analyzing many blog pages from various sites, 
we find that root nodes of all body subtrees contain 
$<$div$>$ or $<$span$>$ or $<$p$>$ tags. So we choose those subtrees as body subtree candidates. 

\subsection{SVM Classifier}
Support Vector Machines(SVM) \cite{burges1998tutorial} are famous supervised learning models, 
with associated learning algorithms that analyze data and recognize patterns. 
We can use it for the purpose of classification and regression analysis. 
Using kernel tricks, SVM can perform not only linear classification but also non-linear classification. 

We first represent every node in DOM-Tree by a feature vector and then 
build two distinguished SVM classifiers with Gaussian RBF kernel \cite{chapelle1999support}
to extract titles and bodies from blog pages respectively. 
Features used to learn SVM classifiers will be described with details in the next section.

\section{Feature Construction}
\subsection{Title Subtree}
People can recognize the title at first glance because of its outstanding position 
and size. According to this fact, spatial features in titles can be exploited to identify the set $S$ of
title subtrees. Moreover, some content features are also very helpful to identify title subtrees. 
Spatial and content features used in title extraction include:
\subsubsection{Spatial features.}  
\begin{itemize}
\item $TRectCenter_N$. Let \textit{TRect} denote the bounding rectangle of a subtree $\mathbf{t}_i$ and  
\textit{TRectCenter} denote the displaying center coordinates $(x, y)$ of \textit{TRect} in the browser. 
As different blog sites having different layouts, 
we need to normalize \textit{TRectCenter} with \textit{BrowserCenter}, 
which is the center coordinates $(x, y)$ of the browser: 
\begin{equation}
TRectCenter_{N}=\frac{TRectCenter-BrowserCenter}{||BrowserCenter||}
\end{equation}
\item$FontSize$. \textit{FontSize} is about the font size of words in a subtree ${\bf t}_i$. 
It helps us to distinguish title subtrees from
other subtrees in two aspects. One is that titles are usually
displaying in font sizes large enough to be differentiated from
other contents. The other one is that in most blog pages, all titles are
displaying in the same font size. 
\end{itemize}

\subsubsection{Content features.} 
\begin{itemize}
\item$TitleLen$. \textit{TitleLen} is the length of texts measured by 
the number of words contained in a subtree ${\bf t}_i$. 
This feature helps us to distinguish titles from long passages, 
based on the fact that the title seldom has too many words in it. 
\item$Link$. \textit{Link} is the state about the hyperlink in a subtree $\mathbf{t}_i$. 
There are three situations for this feature: \textit{no 
link}, \textit{internal} and \textit{external}. 
Because the title usually contains an internal link to its corresponding post page, 
so the state of \textit{Link} is usually \textit{internal}, 
sometimes \textit{no link} and seldom \textit{external}.
\item$EndWith$. \textit{EndWith}
describes whether the text contained in a subtree $\mathbf{t}_i$ ends up with a colon, or
a semicolon or a full stop. Our observation is that a title seldom
ends with these punctuation marks.
\end{itemize}
\subsection{Body Subtree}
Bodies in a blog page consist of discrete passages separated by
titles. The extraction of them poses great challenges in that their contents may vary 
greatly (e.g. no texts, only pictures) and they are displayed in styles of great diversity. 
There are several title blocks and body blocks in a single blog page and 
they are in alternating arrangement, which can be seen from Fig.\ref{fig:blog-home-page} easily. 
We can use titles having been extracted to identify bodies better. 
Spatial and
content features used in body extraction include:

\subsubsection{Spatial features.}
\begin{itemize}
\item$Widthper$. \textit{Widthper} is the width of a subtree $\mathbf{t}_i$ in percentage of the screen width. 
This feature helps us to eliminate many unrelated contents such as \textit{publishing time}, \textit{clicked times}, 
which are usually of short width. 
\item$RelV$. \textit{RelV} is about relative vertical positions of a subtree ${\bf t}_i$ compared to extracted title blocks. 
This feature helps distinguish the body block from other blocks at the top of 
the blog page, such as introduction of the blogger, which are too hard to eliminate if we only 
consider the content features. 
\item$RelH$. \textit{RelH} is about relative horizontal positions of a subtree ${\bf t}_i$ with respect to 
extracted title blocks. According to our observation, 
the title block and the body block are close in the horizontal position. 
Moreover, the central of the title block always lies between the left and right boundary of the body block, 
as shown in Fig.\ref{fig:title-body-relation}. 
This feature can help us to eliminate noises located far from the title, such as Ads, related links etc. 
\begin{figure}[ht]
\centering
\includegraphics[height=3.5cm]{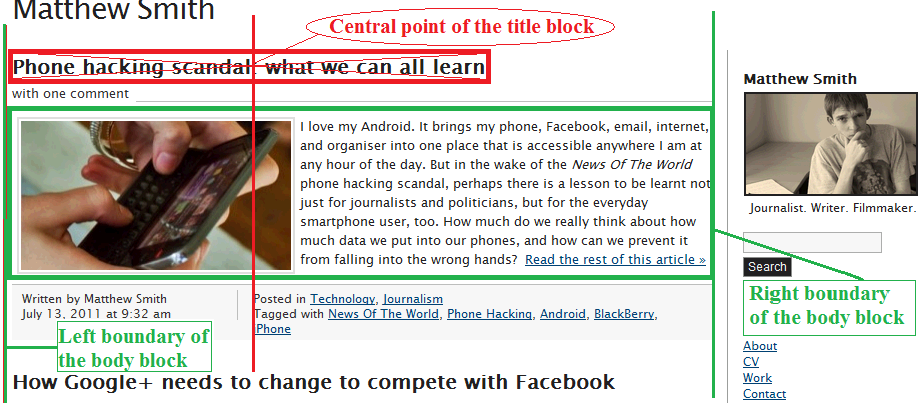}
\caption{\label{fig:title-body-relation}
An example showing the relation between the title block and body block.}
\end{figure}
\item $BRectCenter_N$. Let \textit{BRect} denote the bounding rectangle of a subtree $\mathbf{t}_i$ and 
\textit{BRectCenter} denote the center coordinates $(x, y)$ of \textit{BRect} displaying in the browser. 
Like the \textit{TRectCenter}, \textit{BRectCenter} also need to be normalized using \textit{BrowserCenter},
which has been mentioned in the context of title subtree features: 
\end{itemize}
\begin{equation}
BRectCenter_{N}=\frac{BRectCenter-BrowserCenter}{||BrowserCenter||}
\end{equation}
\subsubsection{Content features.}  
{
\begin{itemize}
\item$BodyLen$. \textit{BodyLen} is the number of words contained in a subtree $\mathbf{t}_i$. 
This feature is based on the fact that 
the body block often contains more words than other blocks.
\item$MarksNum$. \textit{MarksNum} is the 
number of punctuation marks in a subtree $\mathbf{t}_i$. 
Bodies often contain more punctuation marks,
so this feature helps us to distinguish the body block from other blocks such as \textit{Ads}, \textit{related links} etc.  
\item$EndWith$. \textit{EndWith} describes whether the text of a subtree $\mathbf{t}_i$
ends with an ellipsis (...). The bodies listed in the blog page are usually not
complete and end with ellipses.
\end{itemize}
\subsection{Standardization}
We have used the RBF kernel function for SVM, so all features should be centered around $0$ and 
have variance in the same order by removing the mean and scaling to unit variance. 
What's more, it can also reduce the time to find support vectors in SVM \cite{tax2000feature}. 
\section{Experiments and Results}
\subsection{Experimental Setup}
In our experiments, we crawled 2,250 blog pages from nine very different blog sites and labeled them manually. 
More precisely, 250 page are crawled and labeled for each site. 
The blog sites from which we crawled blog pages are listed in Table \ref{tab:blog-site}. 
\begin{table}[ht!]
\centering \caption{
Blog sites used in our experiments} \label{tab:blog-site}
\begin{tabular}{p{3cm} p{3.5cm}} \hline

{\bf Blog Site}&{\bf URL}\\
\hline
Blogger          & www.blogger.com \\
Wordpress        & www.wordpress.com \\
Planet Eclipse    & www.planeteclipse.com \\
Boingboing       & www.boingboing.net \\
Myspace         & www.myspace.com \\
Msdn             & blogs.msdn.com \\
Yahho            & blog.yahoo.com \\
Sina             & blog.sina.com.cn \\
Qzone            & blog.qq.com \\[2pt]
\hline\end{tabular}
\end{table}

For all experiments in the following sections, the results are averaged over multiple runs of the experiments.
\subsection{Single Blog Site Experiment}
In this experiment, we aim to learn the extraction accuracy of our method by training and testing the extraction model on a single blog site. 
For each blog site, we used 10, 20, 30, 40 blog pages as training set and trained SVM classifiers respectively. 
Then we tested each SVM classifier on the remaining blog pages of corresponding site. 
The results are shown in Table \ref{tab:single-blog-accuracy}. 
Note that the accuracy in this table is considering the title and the body together. 
\begin{table}[ht!]
\centering \caption{
Extraction accuracy for classifiers trained and tested on a single blog site} \label{tab:single-blog-accuracy} 
\begin{tabular}{p{2cm} p{1cm} p{1cm} p{1cm} p{1cm} } \hline
 {\bf Blog Site}&{\bf 10}&{\bf 20}&{\bf 30}&{\bf 40}\\
\hline Blogger & 99.3$\%$ & 100$\%$ & 99.3$\%$ & 100$\%$ \\
Wordpress & 96.7$\%$ & 98.7$\%$ & 98.7$\%$ & 98.7$\%$  \\
Planet Eclipse & 100$\%$ & 100$\%$ & 100$\%$ & 100$\%$  \\
Boingboing & 98.7$\%$ & 98.7$\%$ & 98.7$\%$ & 98.7$\%$  \\
Myspace & 99.3$\%$ & 99.3$\%$ & 99.3$\%$ & 99.3$\%$  \\
Msdn & 100$\%$ & 100$\%$ & 100$\%$ & 100$\%$  \\
Yahho & 96.7$\%$ & 100$\%$ & 98.7$\%$ & 98.7$\%$  \\
Sina & 96.7$\%$ & 96.7$\%$ & 96.7$\%$ & 98.7$\%$  \\
Qzone & 99.3$\%$ & 98.7$\%$ & 98.7$\%$ & 98.7$\%$  \\
\hline {\bf Average} & {\bf 98.5}$\%$ & {\bf 99.1}$\%$ & {\bf 98.9}$\%$& {\bf 99.2}$\%$\\
\hline\end{tabular}
\end{table}

The results show that our method achieves a high accuracy for extracting titles and bodies in a single blog site. 
We can also see from the table that the number of training samples  
has little effect in the extraction accuracy. The reason for this high accuracy using only a small number of training pages is that the layout and structure for pages in a single blog site tend to be consistent. 
\\
\subsection{Multiple Blog Sites Experiment}
In this experiment, we used 360 blog pages from 9 blog sites (40 blog pages from each blog site) to construct the training set. 
The trained SVM classifiers were then tested on the remaining blog pages of corresponding site. 
The experimental results are shown in Table \ref{tab:all-blog-accuracy}.
\begin{table}[ht!]
\centering \caption{
Extraction accuracy for classifiers trained and tested on multiple blog sites} \label{tab:all-blog-accuracy} 
\begin{tabular}{p{3cm} p{1.5cm} p{1.5cm} } \hline

{\bf Blog Site}&{\bf Title}&{\bf Body}\\
\hline Blogger          & 100$\%$ & 96.7$\%$ \\
WordPress        & 98.7$\%$ & 98.7$\%$ \\
Plant Eclipse    & 100$\%$ & 98.7$\%$ \\
Boingboing       & 98.7$\%$ & 95.3$\%$ \\
Myspace          & 99.3$\%$ & 99.3$\%$ \\
Msdn             & 100$\%$ & 100$\%$ \\
Yahho            & 100$\%$ & 100$\%$ \\
Sina             & 96.7$\%$ & 93.3$\%$ \\
Qzone            & 98.7$\%$ & 96.7$\%$ \\
\hline {\bf Average} & {\bf 99.1}$\%$ & {\bf 97.6}$\%$\\
\hline\end{tabular}
\end{table}

Though some drop in the extraction accuracy can be observed in this experiment compared with the previous one on a single blog site, the extraction accuracy in general is still satisfactory. This experiment shows our method has the capability to accommodate the difference in page layout and structure, which is an important characteristic for template-independent extractors.
From the experimental results in table \ref{tab:all-blog-accuracy}, we can also see  that body extraction 
is more difficult than title extraction.

\subsection{Extraction Accuracy under Varying Training Set}
This experiment was performed with the number of training pages  from
each site varying from 2 to 20 with a step of 2.
Fig.\ref{fig:change} shows the change of average
accuracy for title and body extraction. 
\begin{figure}[ht!]
\centering
\includegraphics[height=6cm]{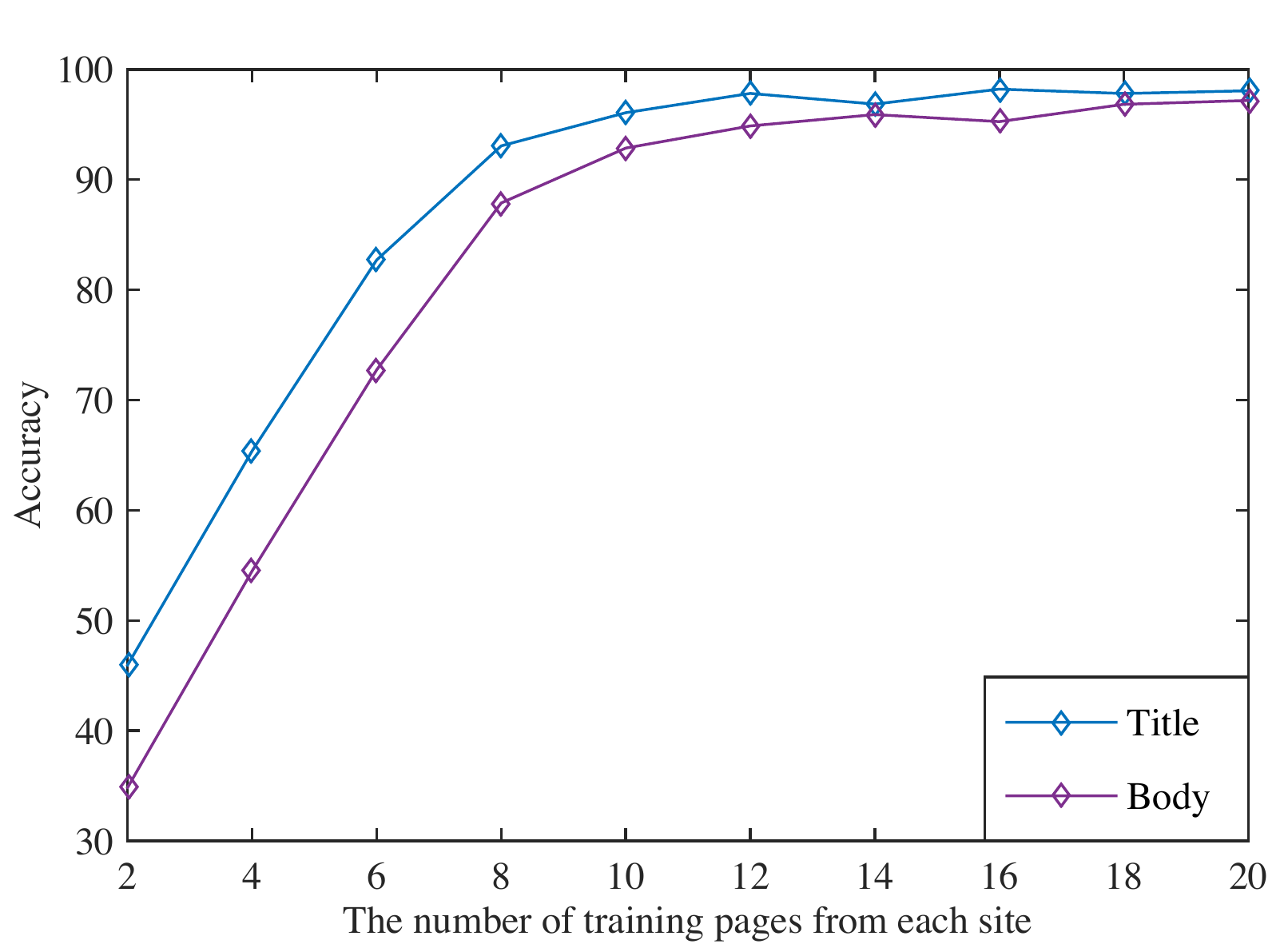}
\caption{\label{fig:change}
The change of accuracy over the number of training pages.}
\end{figure}

From the experimental results, we can see the extraction accuracy for
both titles and bodies improves when the number of training pages
increases. When more than $10$ training pages for each blog site are used, the
extraction accuracy is stabilized at about 97\% and 95\% for title
and body respectively.
\\
\subsection{Generalization Capability of the Extractor}
In this experiment, we tested the generalization capability of our method. 
We randomly chose 100 blog pages from all blog pages as the training set and the rest 2,150 blog pages as  the testing set.  
The results are shown in Table \ref{tab:generalization-accuracy}. 
\begin{table}[ht!]
\centering \caption{Extraction accuracy for classifiers trained on randomly selected pages, which shows their generalizatidon capability} \label{tab:generalization-accuracy}
\begin{tabular}{p{3cm}p{1.5cm}p{1.5cm}} \hline

{\bf Blog Site}&{\bf Title}&{\bf Body}\\
\hline Blogger          & 89.3$\%$ & 87.5$\%$ \\
WordPress        & 88.6$\%$ & 87.1$\%$ \\
Plant Eclipse    & 91.3$\%$ & 89.8$\%$ \\
Boingboing       & 90.3$\%$ & 87.9$\%$ \\
Myspace          & 91.6$\%$ & 89.3$\%$ \\
Msdn             & 92.3$\%$ & 90.8$\%$ \\
Yahho            & 89.6$\%$ & 88.7$\%$ \\
Sina             & 83.8$\%$ & 80.2$\%$ \\
Qzone            & 80.2$\%$ & 78.8$\%$ \\
\hline{\bf Average }&{\bf 88.6}$\%$ & {\bf 86.7}$\%$\\
\hline\end{tabular}
\end{table}

As can be seen, the extractor achieves an average accuracy 88.6$\%$ and 86.7$\%$ for title and body extraction respectively, 
which demonstrates that our method has a good generalization capability.

\section{Conclusion and Future Work}
In this paper, we present a novel and effective method to construct 
template-independent content extractor for blog pages. 
This extractor is learned from a set of blog pages automatically. 
After parsing a blog page into a DOM-Tree, 
the subtree candidate set is constructed for titles and bodies respectively. 
Then we train SVM classifiers using both spatial and content features. 
Moreover, we use the relation between body blocks and title blocks when we deal with body extraction. 
Finally, experimental results show that our template-independent extractor have a high accuracy 
and it is very stable when dealing with different blog sites. 

As future work, we first consider to develop methods to construct better subtree candidate sets. 
Then we consider to find more efficient features as the input of SVM classifiers, 
in order to improve the generalization of our method. 
What's more, we will also try to apply the whole framework to other applications, 
such as microblog extraction. 

\section*{Acknowledgment}
This work is supported by Zhejiang Provincial Natural Science Foundation of China(Grant no.LZ13F020001), 
National Science Foundation of China(Grant nos.61173185,61173186). 

\bibliographystyle{IEEEtran}
\bibliography{ref}
\end{document}